\documentclass[aip,reprint,amsmath,amssymb,floatfix,superscriptaddress]{revtex4-1}
\usepackage[colorlinks, citecolor=blue, linkcolor=blue]{hyperref}
\usepackage{graphicx}
\usepackage{bm}
\usepackage{txfonts}

\begin{document}

\title{A magnetic domain wall Mackey-Glass oscillator}

\author{J{\'e}r{\^o}me Williame}
\author{Joo-Von Kim}
\email{joo-von.kim@c2n.upsaclay.fr}
\affiliation{Centre de Nanosciences et de Nanotechnologies, CNRS, Universit\'e Paris-Saclay, 91120 Palaiseau, France}

\date{\today}

\begin{abstract}
We propose a time-delay oscillator with Mackey-Glass nonlinearity based on a pinned magnetic domain wall in a thin film nanostrip. Through spin transfer torques, electric currents applied along the strip cause the domain wall to deform and displace away from a geometrical pinning site, which can be converted into a nonlinear transfer function through a suitable choice of a readout. This readout serves as a delay signal, which is subsequently fed back into the applied current with amplification. With micromagnetics simulations, we study the role of the readout position, time delay, and feedback gain on the dynamics of this domain wall. In particular, we highlight regimes in which self-sustained oscillations and complex transients are possible.
\end{abstract}

\maketitle

Magnetic domain walls in thin films possess a number of useful properties for information technologies. Domain walls are topological solitons whose spatial profile is determined by competing exchange interactions and anisotropies, which confer them particle-like properties that make them useful for binary storage or logic gates. For example, the possibility to move sequences of domain walls back and forth under applied currents~\cite{Grollier:2003kt, Vernier:2004dg, Yamaguchi:2004in, Yamanouchi:2004cs, Klaui:2005hb} allows for information to be shuttled back and forth in magnetic shift registers~\cite{Parkin:2008gs}. Domain wall devices have been touted as a promising avenue to reduce the energy consumption in information technologies.

Beyond applications in storage and logic, attention in spintronics has recently turned towards other forms of non-Boolean information processing such as neuro-inspired computing. Such efforts are motivated by the goal of identifying ways to mimic the capacity of the human brain to efficiently perform cognitive tasks, such as pattern recognition, with noisy or incomplete data. The inherent nonlinearity in magnetization dynamics is useful in this respect, since neuronal dynamics exhibit complex transients, self-oscillations, and chaotic behavior. For example, it has been shown recently that spin-torque nano-oscillators can be used for classification tasks~\cite{Torrejon:2017hj, Romera:2018dm, Riou:2019ik}.

Here, we examine a feature of neuronal dynamics that has been explored sporadically in spintronics, namely that of magnetization dynamics with delayed feedback effects. In general terms, this entails feeding the output of the dynamical system, after amplification and a time delay, back into its input as an additional component of the external drive (e.g., fields or  currents). Delayed feedback has been explored in terms of self-injection in spin-torque nano-oscillators, where their sensitivity to external perturbations~\cite{Tiberkevich:2014eh} and changes to characteristics such as the spectral linewidth have been examined~\cite{Khalsa:2015kn, Tamaru:2016kl, Tsunegi:2016ka, Singh:2017gt, Singh:2018gr, williame:2020eo}. Feedback in low-dimensional systems such as macrospin models can result, for example, in nontrivial behavior such as temporal chaos~\cite{Williame:2019hd, Taniguchi:2019ej}. In other systems, the onset of self-sustained oscillations can appear through a variety of different feedback mechanisms~\cite{Dixit:2012ec, Kumar:2016io, Bhuktare:2017ja}. More broadly, delayed-feedback has been employed in nonlinear elements as building blocks of time-delay architecture for reservoir computing~\cite{Appeltant:2011jy, Larger:2017cx}, a neuro-inspired scheme that has garnered growing interest in magnetism and spintronics~\cite{Bourianoff:2018gp, Prychynenko:2018ii, Nakane:2018jc, Markovic:2019dz, Riou:2019ik, Araujo:2020ro, Yamaguchi:2020hg, Pinna:2020rc}.

A prototypical example of delayed feedback dynamics is the Mackey-Glass oscillator, which is described by the one-dimensional delay differential equation~\cite{Mackey:1977dv},
\begin{equation}
\dot{x}(t) = -\Gamma x(t) + \frac{C x(t-\tau)}{1+x(t-\tau)^p}.
\label{eq:MG}
\end{equation}
The right-hand side of the equation comprises two terms. The first represents damping with a relaxation rate $\Gamma$. The second is the delayed feedback term, where $C$ is an amplification, and $p$ determines the nonlinearity of the transfer function. Depending on the values of $\Gamma$, $C$, and $p$, the system in Eq. (\ref{eq:MG}) can exhibit complex transients, self-oscillations, and chaos. In the absence of delayed feedback, Eq.~(\ref{eq:MG}) exhibits only trivial relaxation dynamics. Optoelectronic schemes exploiting the kind of Mackey-Glass nonlinearity in Eq.~(\ref{eq:MG}) have been recently used in time-delay architectures for reservoir computing~\cite{Appeltant:2011jy, Larger:2017cx}. In the context of neuro-inspired information processing, it is, therefore, interesting to enquire how similar implementations could be realized using magnetic states and spintronic systems.

\begin{figure}
\centering\includegraphics[width=7.5cm]{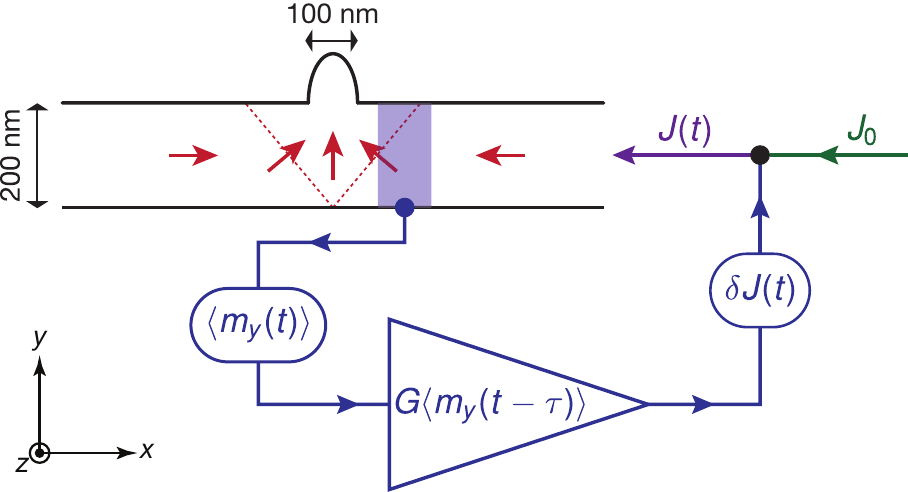}
\caption{Geometry of the domain wall Mackey-Glass oscillator. The magnetic strip, 200 nm in width, hosts an in-plane magnetized transverse domain wall, which is pinned by an elliptical protrusion of width 100 nm. A dc conventional current $J_0$ flows along the track and displaces the wall through spin-transfer torques. A magnetoresistance readout sensor detects the $\langle m_y \rangle$ component near the protrusion, which is amplified by a factor $G$ and fed back into the input current as a delayed modulation.}
\label{fig:Geometry}
\end{figure}

Our proposal for a domain wall Mackey-Glass oscillator (DWMGO) is presented in Fig.~\ref{fig:Geometry}.
%
The system comprises an in-plane magnetized ferromagnetic strip with an elliptical protrusion, which acts as a strong pinning site for a transverse domain wall, similar to a magnetic racetrack memory~\cite{Parkin:2008gs}. An electric current is applied along $x$, the long axis of the strip, which simultaneously leads to a displacement and a distortion of the domain wall along this direction. A magnetoresistive readout sensor is positioned next to the protrusion, which is sensitive to the $y$ component of the magnetization. The signal from this readout, which represents the spatial average across the sensor at position $i$, $\langle m_y(t) \rangle_i$, is amplified by a factor $G$ with a time delay $\tau$ and subsequently fed back into the input as an additional time-dependent current, $\delta J(t)$. The current-driven domain wall dynamics is then subjected to the time-dependent drive,
\begin{equation}
J(t) = J_0 + G a_i \, \langle m_y(t-\tau) \rangle_i,
\label{eq:curr}
\end{equation}
where $a_i$ is a normalization constant discussed below.

The dynamics of the DWMGO is simulated using the \textsc{MuMax3} code~\cite{Vansteenkiste:2014et}, which performs a numerical time integration of the Landau-Lifshitz equation with current-induced spin torques,
\begin{equation}
\frac{d \mathbf{m}}{dt} = -\gamma_0 \mathbf{m} \times \mathbf{H}_\mathrm{eff} + \alpha \mathbf{m} \times \frac{d \mathbf{m}}{dt} - \left(\mathbf{u}\cdot\mathbf{\nabla}\right)\mathbf{m}+ \beta \mathbf{m} \times \left(\mathbf{u}\cdot\mathbf{\nabla}\right)\mathbf{m}.
\label{eq:LLG}
\end{equation}
$\gamma_0$ is the gyromagnetic ratio, $\mathbf{m}(\mathbf{r},t)$ is a unit vector representing the magnetization field, $\mathbf{H}_\mathrm{eff}$ is the effective magnetic field, and $\alpha$ is the Gilbert damping constant. The Zhang-Li spin transfer torques are parametrized by $\mathbf{u} = \mathbf{J}(\mathbf{r}) P \mu_B/ (e M_s)$, which represents an effective spin drift velocity where $\mathbf{J}$ is the conventional current density, $\mu_B$ is the Bohr magneton, $e$ is the electron charge, $\beta$ is the nonadiabaticity, and $P$ is the spin polarization. The effective field is given by the variational derivative of the total magnetic energy $U$ with respect to the magnetization unit vector, $\mathbf{H}_\mathrm{eff} = - (1/\mu_0 M_s) \delta U/\delta \mathbf{m}$, and comprises contributions from the exchange, dipole-dipole, and the Zeeman interactions.

\begin{figure}
\centering\includegraphics[width=7cm]{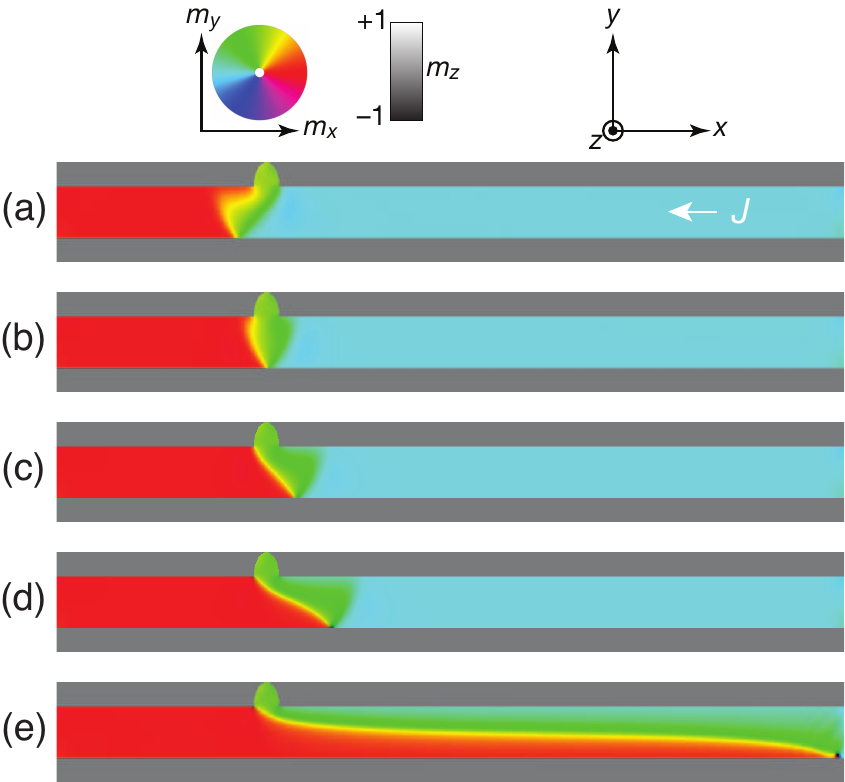}
\caption{Static domain wall profiles under different applied dc currents, $J(t) = J_0$. (a) $J_0 = -4$, (b) $J_0 = 0$, (c) $J_0 = 4$, (d) $J_0 = 8$, and (e) $J_0 = 10$ TA/m$^2$. The critical depinning current is approximately $J_{0,\mathrm{c}}$ = 8.5 TA/m$^2$. The functionality of the DWMGO is limited to currents below this critical current.}
\label{fig:DW_config}
\end{figure}

The oscillator studied comprises a strip with dimensions of $3072 \times 200 \times 10$ nm, which is discretized using finite difference cells $3 \times 3 \times 10$ nm in size. The protrusion is taken to be a half ellipse with the long axis perpendicular to the wire, which extends out to a distance of 92 nm from the wire, and the short axis parallel to the wire with an extension of 100 nm. We assume micromagnetic parameters relevant for permalloy, namely a saturation magnetization of $M_s = 860$ kA/m, an exchange constant of $A = 13$ pJ/m, and a Gilbert damping constant of $\alpha = 0.02$. The readout region is taken to be rectangular in shape with dimensions of $100 \times 200$ nm, which spans the entire with of the magnetic strip. For the current-driven spin torques, we assume $P = 1$ for simplicity and a nonadiabacity of $\beta = 0.1$. Finite element electromagnetics simulations were used to compute the spatial profile of the currents flowing along the wire. Current densities used here refer to the average current density applied through the rectangular cross section at the ends of the wire, far from the protrusion. In Fig.~\ref{fig:DW_config}, the static domain wall profile under different applied dc currents is shown.
%

\begin{figure}
\centering\includegraphics[width=7cm]{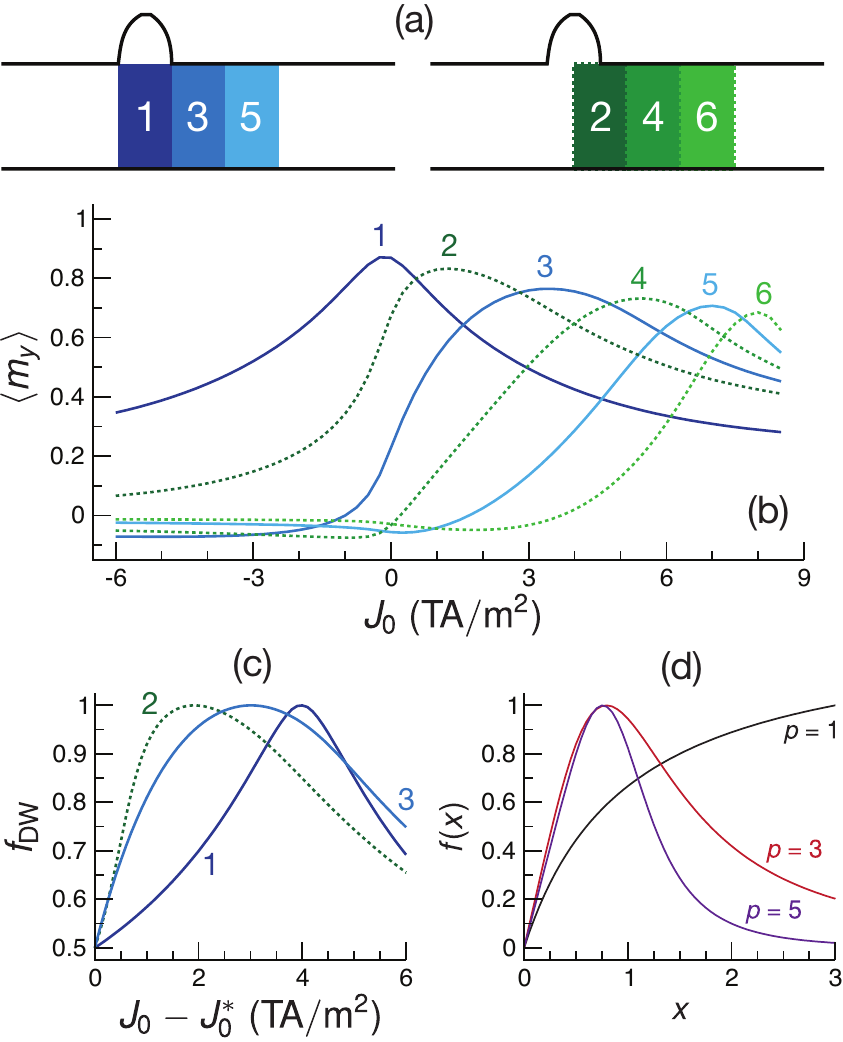}
\caption{Magnetoresistive readout signal of the DWMGO. (a) Example of six sensor positions. (b) Spatial average of $m_y$ as a function of the static applied current $J_0$ for each sensor position in (a). (c) Rescaled output function at the operating point of sensor positions 1 to 3. (d) Mackey-Glass transfer function for three different $p$ for comparison.}
\label{fig:MGfn}
\end{figure}

The magnetoresistive signal, which serves as the transfer function $f$ [i.e., the second term on the right hand side of Eq.~(\ref{eq:MG})], is illustrated in Fig.~\ref{fig:MGfn}.
%
The signal is computed from the spatial average of the magnetization component $m_y$ within the sensor region. Six different positions $i$ of the sensor are shown, which results in different asymmetries and nonlinearities for the transfer function [Fig.~\ref{fig:MGfn}(a)]. To facilitate comparison between the different sensor positions, we rescale the transfer functions in the following way. First, we normalize each curve by a factor $a_i$ such that the maximum value of the output is unity; we define the transfer function to be $f_\mathrm{DW} = a_i \langle  m_y \rangle_i$. Second, we define an operating point $J_{0,i}^*$ such that the value of the transfer function is 0.5 at $J_0 - J_0^* = 0$. The values of $a$ and $J_0$ for each sensor position is given in Table~\ref{tab:params}.
\begin{table}
\caption{\label{tab:params}Scaling parameter $a_i$ and operating point $J_{0,i}^*$ for each sensor position $i$ such that $f_\mathrm{DW}=0.5$ at $J_0=J_{0,i}^*$.}
\begin{ruledtabular}
\begin{tabular}{ccc}
Position, $i$ & $a_i$ & $J_{0,i}^*$ (TA/m$^2$)\\
1 & 1.145 & $-4.12$ \\
2 & 1.201 & $-0.68$ \\
3 & 1.307 & $0.42$ \\
4 & 1.366 & $2.20$ \\
5 & 1.413 & $4.38$ \\
6 & 1.366 & $6.24$ 
\end{tabular}
\end{ruledtabular}
\end{table}
The transfer function for positions 1 to 3 is shown in Fig.~\ref{fig:MGfn}(b). For the purposes of comparison, the transfer function for the Mackey-Glass oscillator [Eq.~(\ref{eq:MG})] is shown in Fig.~\ref{fig:MGfn}(c) for three values of the nonlinearity, $p$. $f_\mathrm{DW}$ captures the non-monotonic behavior of the Mackey-Glass transfer function for $p>1$, with the sensor position offering a means to adapt the nonlinearity.

Next, we discuss the possible dynamical states of the DWMGO and the influence of the sensor position. Simulations with the delayed feedback are performed as follows. For a given value of $\tau$ under study, time integration of Eq.~(\ref{eq:LLG}) is first performed over the interval $t \in [-\tau,0]$ without the feedback term in order to generate the delay term $\langle m_y(t-\tau) \rangle$, which is stored in memory. Since we employ a time integration algorithm with an adaptive time step (i.e., primarily the Dormand-Prince method implemented in \textsc{MuMax3}~\cite{Vansteenkiste:2014et}), we use cubic interpolation for the function $m_y(t-\tau)$ so that it can be evaluated at arbitrary values of $t-\tau$ during the simulation. During this initial phase without feedback, a static current $J_0 = J_0^*$ is applied. The initial magnetization state is computed previously for this operating point in order to minimize transient dynamics. For $t>0$, Eq.~(\ref{eq:LLG}) is solved together self-consistently with Eq.~(\ref{eq:curr}).

\begin{figure}
\centering\includegraphics[width=7cm]{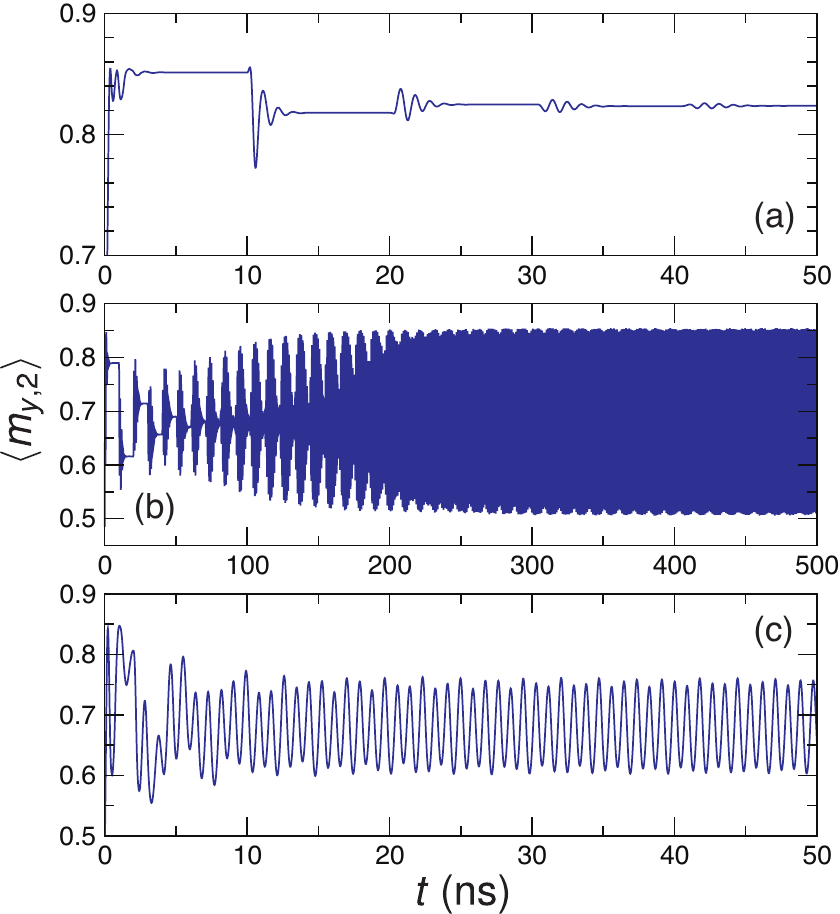}
\caption{Time traces of the magnetization dynamics measured at Position 2. (a) Transient dynamics toward a static configuration under weak amplification, $G = 2$, and long delays, $\tau = 10$ ns. (b) Self-sustained oscillations at $G = 6$ and long delays, $\tau = 10$ ns. Self-sustained oscillations at $G = 6$ and short delays, $\tau = 1$ ns. }
\label{fig:timetrace}
\end{figure}

Some examples of the domain wall dynamics are shown in Fig.~\ref{fig:timetrace}.
%
The figure shows the simulated magnetoresistance readout for a sensor in position 2. It is instructive to first consider the effect of long delays ($\tau = 10$ ns). At low gain, $G = 2$, the feedback acts like a driving term formed by a sequence of steps, where each step corresponds to a different equilibrium position of the domain wall [Fig.~\ref{fig:timetrace}(a)]. As the step is initially applied, a ringdown of the signal can be seen over a few ns, which corresponds to the damped oscillatory motion of the domain wall about the new equilibrium state. Since $f_\mathrm{DW}$ is designed to be finite at $J_0 = J_0^*$, the application of the feedback guarantees some transient behavior since $J$ will deviate from $J_0^*$ instantaneously. The stepped ringdown response eventually converges to a steady state value. At higher gain, $G = 6$, the initial stepped response comprises larger amplitude motion, which over the course of a few hundred ns results in a steady state oscillation of the domain wall, as shown in Fig.~\ref{fig:timetrace}(b). The initial phase is dominated by transients on a time scale give by $\tau$, but this gradually evolves into a self-oscillatory mode for which this modulation is less present. At short delays ($\tau = 1$ ns), the transient dynamics toward the self-oscillation is considerably shorter, which can be observed in Fig.~\ref{fig:timetrace}(c).

\begin{figure}
\centering\includegraphics[width=8.0cm]{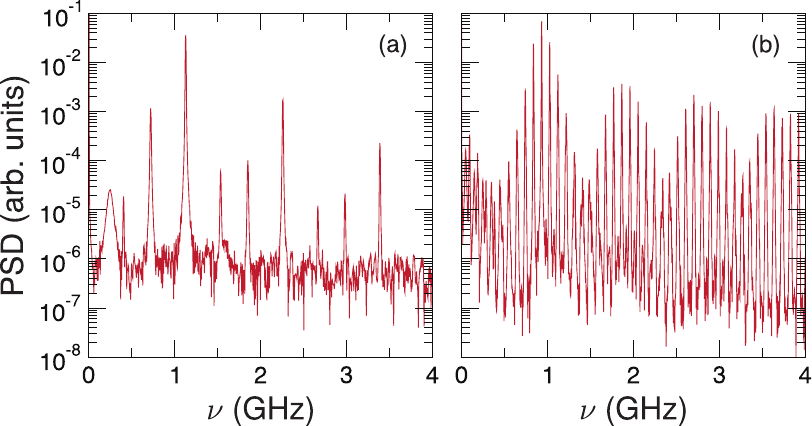}
\caption{Power spectral density (PSD) of domain wall oscillations corresponding to readout at Position 2 with $G=6$ and two different delays: (a) $\tau = 1$ ns and (b) $\tau = 10$ ns.}
\label{fig:psd}
\end{figure}

Examples of the resulting power spectrum of feedback-driven oscillations are presented in Fig.~\ref{fig:psd}.
%
The spectra in Fig.~\ref{fig:psd}(a) and \ref{fig:psd}(b) correspond to the oscillations shown in Fig.~\ref{fig:timetrace}(c) and \ref{fig:timetrace}(b), respectively. Fig.~\ref{fig:psd}(a) presents an example at short delays ($\tau = 1$ ns), where we can clearly identify the main oscillation peak $f_0$ at around 1.13 GHz, with harmonics at $2f_0$ and $3f_0$ also visible in the spectrum shown. These main spectral lines are accompanied by a number of modulation sidebands, which result from a self-modulation that is induced by the delayed feedback signal. This is a feature seen in other delayed-feedback systems involving spintronic devices~\cite{Singh:2017gt, Singh:2018gr, williame:2020eo}.

\begin{figure}
\centering\includegraphics[width=8.5cm]{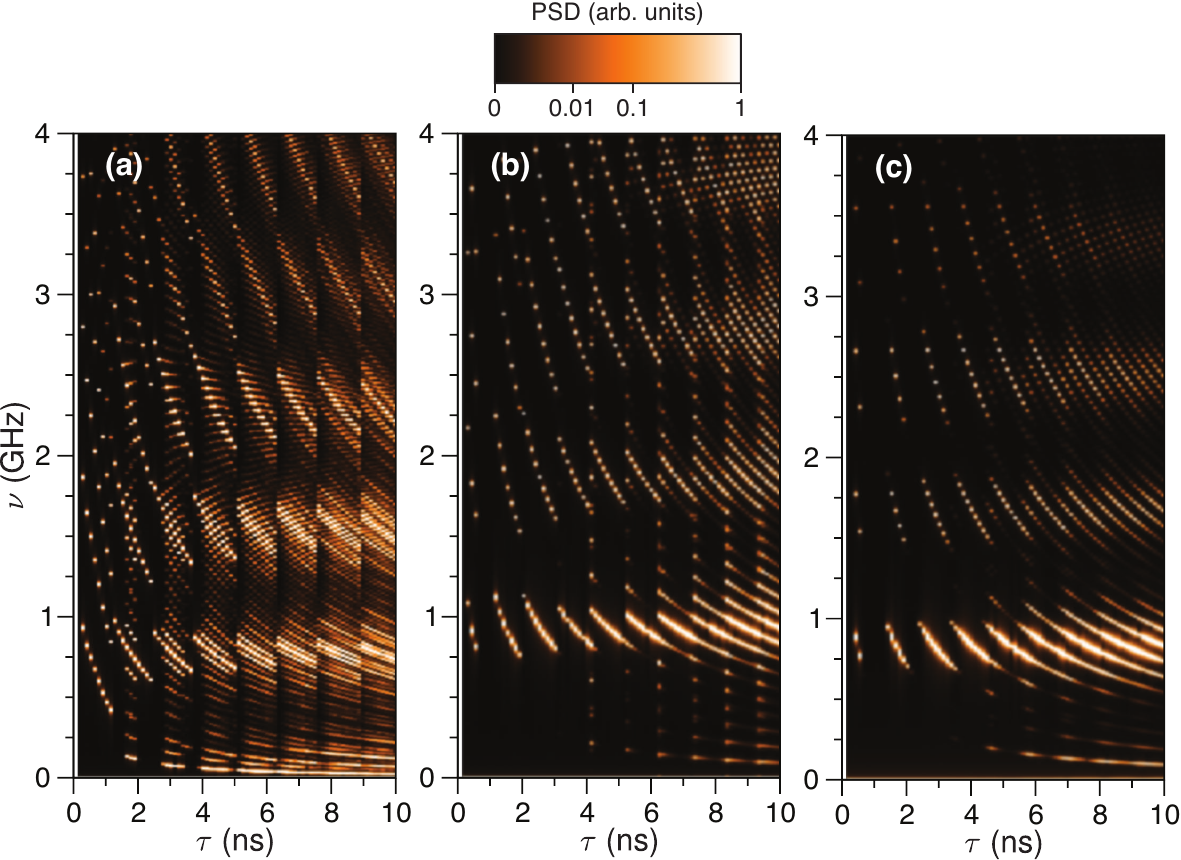}
\caption{Color map of the power spectral density (PSD) as a function of delay $\tau$ for $G = 6$, with the sensor at (a) position 1, (b) position 2, (c) position 3.}
\label{fig:psdtau}
\end{figure}

Figure~\ref{fig:psdtau} presents a color map of the power spectral density of the DWMGO as a function of time delay at a fixed gain of $G=6$, for three sensor positions.
%
For all cases shown, we can observe the primary oscillation peak around 1 GHz, where the strength and number of modulation sidebands vary with the delay. For short delays, the feedback results only in a modulation of the primary oscillation frequency, where a ratchet-like frequency pulling toward lower frequencies is shown as $\tau$ increases. This phenomenon is particularly visible in Fig.~\ref{fig:psdtau}(c). For longer delays, this frequency pulling is accompanied by the appearance of modulation sidebands, which are also visible in the harmonics within the frequency window considered. The onset of these sidebands depend strongly on the sensor position, and therefore, on the feedback nonlinearity. We can see that the sidebands appear at $\tau \simeq 1.5$ ns for a sensor centered at the protrusion (position 1), while at positions 2 and 3 the onset of the sidebands occur at $\tau \simeq 4.0$ and $\tau \simeq 4.5$ ns, respectively. These results show that the rich power spectra can be obtained within a single device by tuning the readout position of the domain wall oscillations.

\begin{figure}
\centering\includegraphics[width=8.5cm]{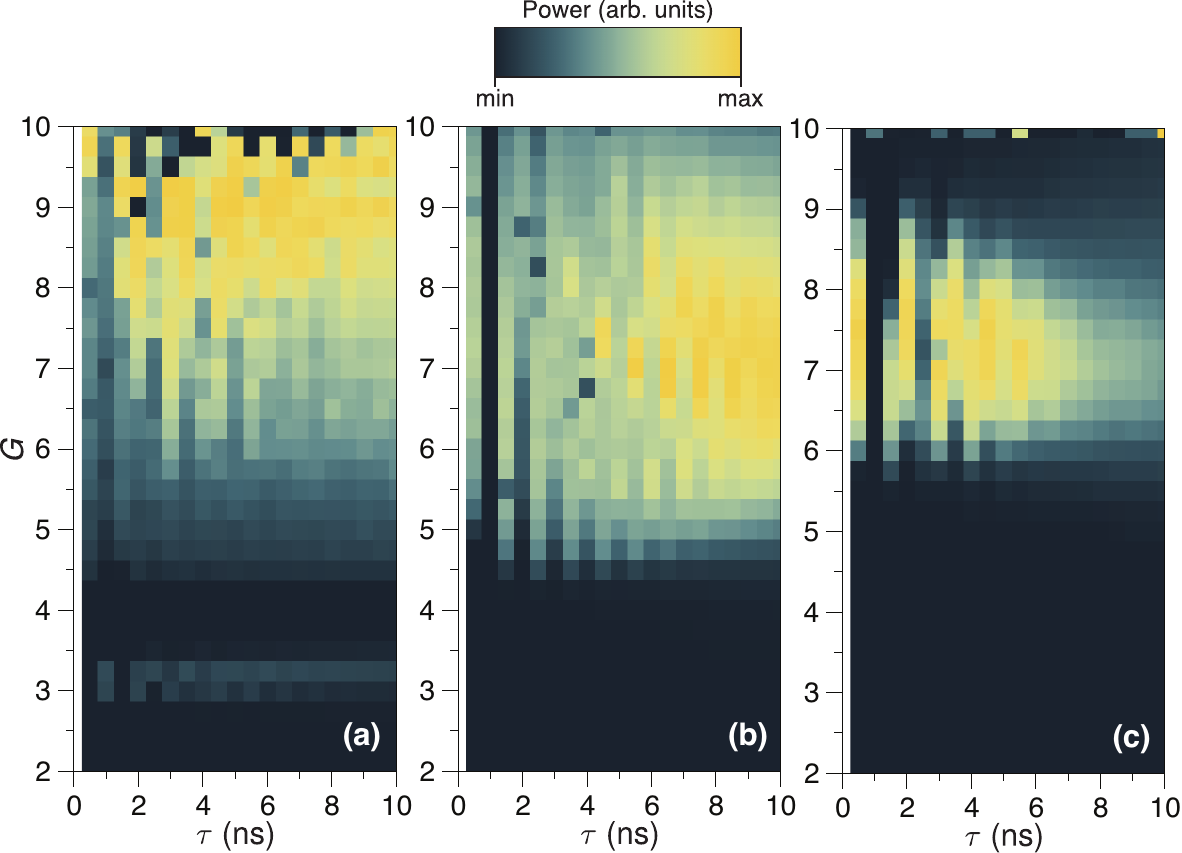}
\caption{Phase diagram of the output power of the oscillator as a function of feedback delay, $\tau$, and gain, $G$, for the readout sensor located at (a) position 1, (b) position 2, and (c) position 3.}
\label{fig:phasediag}
\end{figure}

Figure~\ref{fig:phasediag} presents the output power of the DWMGO for different values of the delay time and amplification with the three sensor positions considered in Fig.~\ref{fig:psdtau}.
%
Similarly to behavior in Fig.~\ref{fig:psdtau}, we can observe modulations in the output power as the delay is varied. Another notable feature is that a maximum in the oscillator power is observed over a range of feedback gain, which can be clearly seen in Fig.~\ref{fig:phasediag}(c). This reflects the fact that the oscillatory properties are strongly dependent on the competition between the relaxation dynamics of the confining potential and the form of the feedback signal.

We now discuss two possible issues and their potential impact on the DWMGO functionality described so far. The first concerns the Joule heating related to the current flow in the device. Because the current densities involved are typical for spin-transfer torques, i.e., in the range of $10^{12}$ A/m$^2$, we can expect the operating temperature of the device to be significantly above ambient conditions. However, since the cross-sectional area at the protrusion is larger than the straight portion of the wire, we do not expect additional complications due to Joule heating near the protrusion or the sensor region, where heating might actually be lower than the baseline value set by the regions far from the protrusion. In terms of thermal noise in the readout, we do not expect any additional issues beyond those faced in designing suitable magnetoresistive readouts, e.g., in three-terminal devices studied in spin-orbit torque switching. The second issue concerns the accuracy of the sensor position, where the vagaries of nanofabrication can also result in variations in its size and shape. Our simulation results using different sensor positions show that the qualitative features of the oscillator remain present irrespective of the sensor position. In a given experimental device, there remain other control parameters such as the feedback delay, feedback gain, and the dc current used as the operating point to tailor the desired dynamics. Shifts in the sensor position away from the position, as shown in Fig.~\ref{fig:MGfn}, result in an overall translation of the nonlinear transfer function along the dc current axis [Fig.~\ref{fig:MGfn}(b)], so variations in the sensor position due to fabrication can be accommodated by tuning this operating point.

In summary, we have presented a model for a spintronic feedback oscillator with a Mackey-Glass nonlinearity. In contrast to conventional spin-torque nano-oscillators in which nonlinearities are determined by intrinsic micromagnetic energies such as shape anisotropies or spin wave interactions, here the form of the nonlinearity is determined in large part by the position of the readout sensor and on how the pinned domain wall deforms as a function of the applied current. This suggests that multiple nonlinearities, and possibly different functionalities, could be designed on a single device by using different readout sensors.

\begin{acknowledgments}
This work was supported by the Agence Nationale de la Recherche (France) under contracts no. ANR-14-CE26-0021 (MEMOS) and ANR-17-CE24-0008 (CHIPMuNCS).
\end{acknowledgments}

\section*{Data availability}
The data that support the findings of this study are available from the corresponding author upon reasonable request.

\bibliography{articles}

\begin{thebibliography}{32}%
\makeatletter
\providecommand \@ifxundefined [1]{%
 \@ifx{#1\undefined}
}%
\providecommand \@ifnum [1]{%
 \ifnum #1\expandafter \@firstoftwo
 \else \expandafter \@secondoftwo
 \fi
}%
\providecommand \@ifx [1]{%
 \ifx #1\expandafter \@firstoftwo
 \else \expandafter \@secondoftwo
 \fi
}%
\providecommand \natexlab [1]{#1}%
\providecommand \enquote  [1]{``#1''}%
\providecommand \bibnamefont  [1]{#1}%
\providecommand \bibfnamefont [1]{#1}%
\providecommand \citenamefont [1]{#1}%
\providecommand \href@noop [0]{\@secondoftwo}%
\providecommand \href [0]{\begingroup \@sanitize@url \@href}%
\providecommand \@href[1]{\@@startlink{#1}\@@href}%
\providecommand \@@href[1]{\endgroup#1\@@endlink}%
\providecommand \@sanitize@url [0]{\catcode `\\12\catcode `\$12\catcode
  `\&12\catcode `\#12\catcode `\^12\catcode `\_12\catcode `\%12\relax}%
\providecommand \@@startlink[1]{}%
\providecommand \@@endlink[0]{}%
\providecommand \url  [0]{\begingroup\@sanitize@url \@url }%
\providecommand \@url [1]{\endgroup\@href {#1}{\urlprefix }}%
\providecommand \urlprefix  [0]{URL }%
\providecommand \Eprint [0]{\href }%
\providecommand \doibase [0]{http://dx.doi.org/}%
\providecommand \selectlanguage [0]{\@gobble}%
\providecommand \bibinfo  [0]{\@secondoftwo}%
\providecommand \bibfield  [0]{\@secondoftwo}%
\providecommand \translation [1]{[#1]}%
\providecommand \BibitemOpen [0]{}%
\providecommand \bibitemStop [0]{}%
\providecommand \bibitemNoStop [0]{.\EOS\space}%
\providecommand \EOS [0]{\spacefactor3000\relax}%
\providecommand \BibitemShut  [1]{\csname bibitem#1\endcsname}%
\let\auto@bib@innerbib\@empty
\bibitem [{\citenamefont {Grollier}\ \emph {et~al.}(2003)\citenamefont
  {Grollier}, \citenamefont {Boulenc}, \citenamefont {Cros}, \citenamefont
  {Hamzic}, \citenamefont {Vaur{\`e}s}, \citenamefont {Fert},\ and\
  \citenamefont {Faini}}]{Grollier:2003kt}%
  \BibitemOpen
  \bibfield  {author} {\bibinfo {author} {\bibfnamefont {J.}~\bibnamefont
  {Grollier}}, \bibinfo {author} {\bibfnamefont {P.}~\bibnamefont {Boulenc}},
  \bibinfo {author} {\bibfnamefont {V.}~\bibnamefont {Cros}}, \bibinfo {author}
  {\bibfnamefont {A.}~\bibnamefont {Hamzic}}, \bibinfo {author} {\bibfnamefont
  {A.}~\bibnamefont {Vaur{\`e}s}}, \bibinfo {author} {\bibfnamefont
  {A.}~\bibnamefont {Fert}}, \ and\ \bibinfo {author} {\bibfnamefont
  {G.}~\bibnamefont {Faini}},\ }\bibfield  {title} {\enquote {\bibinfo {title}
  {{Switching a spin valve back and forth by current-induced domain wall
  motion}},}\ }\href {\doibase 10.1063/1.1594841} {\bibfield  {journal}
  {\bibinfo  {journal} {Applied Physics Letters}\ }\textbf {\bibinfo {volume}
  {83}},\ \bibinfo {pages} {509} (\bibinfo {year} {2003})}\BibitemShut
  {NoStop}%
\bibitem [{\citenamefont {Vernier}\ \emph {et~al.}(2004)\citenamefont
  {Vernier}, \citenamefont {Allwood}, \citenamefont {Atkinson}, \citenamefont
  {Cooke},\ and\ \citenamefont {Cowburn}}]{Vernier:2004dg}%
  \BibitemOpen
  \bibfield  {author} {\bibinfo {author} {\bibfnamefont {N.}~\bibnamefont
  {Vernier}}, \bibinfo {author} {\bibfnamefont {D.~A.}\ \bibnamefont
  {Allwood}}, \bibinfo {author} {\bibfnamefont {D.}~\bibnamefont {Atkinson}},
  \bibinfo {author} {\bibfnamefont {M.~D.}\ \bibnamefont {Cooke}}, \ and\
  \bibinfo {author} {\bibfnamefont {R.~P.}\ \bibnamefont {Cowburn}},\
  }\bibfield  {title} {\enquote {\bibinfo {title} {{Domain wall propagation in
  magnetic nanowires by spin-polarized current injection}},}\ }\href {\doibase
  10.1209/epl/i2003-10112-5} {\bibfield  {journal} {\bibinfo  {journal}
  {Europhysics Letters (EPL)}\ }\textbf {\bibinfo {volume} {65}},\ \bibinfo
  {pages} {526--532} (\bibinfo {year} {2004})}\BibitemShut {NoStop}%
\bibitem [{\citenamefont {Yamaguchi}\ \emph {et~al.}(2004)\citenamefont
  {Yamaguchi}, \citenamefont {Ono}, \citenamefont {Nasu}, \citenamefont
  {Miyake}, \citenamefont {Mibu},\ and\ \citenamefont
  {Shinjo}}]{Yamaguchi:2004in}%
  \BibitemOpen
  \bibfield  {author} {\bibinfo {author} {\bibfnamefont {A.}~\bibnamefont
  {Yamaguchi}}, \bibinfo {author} {\bibfnamefont {T.}~\bibnamefont {Ono}},
  \bibinfo {author} {\bibfnamefont {S.}~\bibnamefont {Nasu}}, \bibinfo {author}
  {\bibfnamefont {K.}~\bibnamefont {Miyake}}, \bibinfo {author} {\bibfnamefont
  {K.}~\bibnamefont {Mibu}}, \ and\ \bibinfo {author} {\bibfnamefont
  {T.}~\bibnamefont {Shinjo}},\ }\bibfield  {title} {\enquote {\bibinfo {title}
  {{Real-Space Observation of Current-Driven Domain Wall Motion in Submicron
  Magnetic Wires}},}\ }\href {\doibase 10.1103/physrevlett.92.077205}
  {\bibfield  {journal} {\bibinfo  {journal} {Physical Review Letters}\
  }\textbf {\bibinfo {volume} {92}},\ \bibinfo {pages} {077205} (\bibinfo
  {year} {2004})}\BibitemShut {NoStop}%
\bibitem [{\citenamefont {Yamanouchi}\ \emph {et~al.}(2004)\citenamefont
  {Yamanouchi}, \citenamefont {Chiba}, \citenamefont {Matsukura},\ and\
  \citenamefont {Ohno}}]{Yamanouchi:2004cs}%
  \BibitemOpen
  \bibfield  {author} {\bibinfo {author} {\bibfnamefont {M.}~\bibnamefont
  {Yamanouchi}}, \bibinfo {author} {\bibfnamefont {D.}~\bibnamefont {Chiba}},
  \bibinfo {author} {\bibfnamefont {F.}~\bibnamefont {Matsukura}}, \ and\
  \bibinfo {author} {\bibfnamefont {H.}~\bibnamefont {Ohno}},\ }\bibfield
  {title} {\enquote {\bibinfo {title} {{Current-induced domain-wall switching
  in a ferromagnetic semiconductor structure}},}\ }\href {\doibase
  10.1038/nature02441} {\bibfield  {journal} {\bibinfo  {journal} {Nature}\
  }\textbf {\bibinfo {volume} {428}},\ \bibinfo {pages} {539--542} (\bibinfo
  {year} {2004})}\BibitemShut {NoStop}%
\bibitem [{\citenamefont {Kl{\"a}ui}\ \emph {et~al.}(2005)\citenamefont
  {Kl{\"a}ui}, \citenamefont {Vaz}, \citenamefont {Bland}, \citenamefont
  {Wernsdorfer}, \citenamefont {Faini}, \citenamefont {Cambril}, \citenamefont
  {Heyderman}, \citenamefont {Nolting},\ and\ \citenamefont
  {R{\"u}diger}}]{Klaui:2005hb}%
  \BibitemOpen
  \bibfield  {author} {\bibinfo {author} {\bibfnamefont {M.}~\bibnamefont
  {Kl{\"a}ui}}, \bibinfo {author} {\bibfnamefont {C.~A.~F.}\ \bibnamefont
  {Vaz}}, \bibinfo {author} {\bibfnamefont {J.~A.~C.}\ \bibnamefont {Bland}},
  \bibinfo {author} {\bibfnamefont {W.}~\bibnamefont {Wernsdorfer}}, \bibinfo
  {author} {\bibfnamefont {G.}~\bibnamefont {Faini}}, \bibinfo {author}
  {\bibfnamefont {E.}~\bibnamefont {Cambril}}, \bibinfo {author} {\bibfnamefont
  {L.~J.}\ \bibnamefont {Heyderman}}, \bibinfo {author} {\bibfnamefont
  {F.}~\bibnamefont {Nolting}}, \ and\ \bibinfo {author} {\bibfnamefont
  {U.}~\bibnamefont {R{\"u}diger}},\ }\bibfield  {title} {\enquote {\bibinfo
  {title} {{Controlled and Reproducible Domain Wall Displacement by Current
  Pulses Injected into Ferromagnetic Ring Structures}},}\ }\href {\doibase
  10.1103/physrevlett.94.106601} {\bibfield  {journal} {\bibinfo  {journal}
  {Physical Review Letters}\ }\textbf {\bibinfo {volume} {94}},\ \bibinfo
  {pages} {106601} (\bibinfo {year} {2005})}\BibitemShut {NoStop}%
\bibitem [{\citenamefont {Parkin}, \citenamefont {Hayashi},\ and\ \citenamefont
  {Thomas}(2008)}]{Parkin:2008gs}%
  \BibitemOpen
  \bibfield  {author} {\bibinfo {author} {\bibfnamefont {S.~S.~P.}\
  \bibnamefont {Parkin}}, \bibinfo {author} {\bibfnamefont {M.}~\bibnamefont
  {Hayashi}}, \ and\ \bibinfo {author} {\bibfnamefont {L.}~\bibnamefont
  {Thomas}},\ }\bibfield  {title} {\enquote {\bibinfo {title} {{Magnetic
  Domain-Wall Racetrack Memory}},}\ }\href {\doibase 10.1126/science.1145799}
  {\bibfield  {journal} {\bibinfo  {journal} {Science}\ }\textbf {\bibinfo
  {volume} {320}},\ \bibinfo {pages} {190--194} (\bibinfo {year}
  {2008})}\BibitemShut {NoStop}%
\bibitem [{\citenamefont {Torrejon}\ \emph {et~al.}(2017)\citenamefont
  {Torrejon}, \citenamefont {Riou}, \citenamefont {Araujo}, \citenamefont
  {Tsunegi}, \citenamefont {Khalsa}, \citenamefont {Querlioz}, \citenamefont
  {Bortolotti}, \citenamefont {Cros}, \citenamefont {Yakushiji}, \citenamefont
  {Fukushima}, \citenamefont {Kubota}, \citenamefont {Yuasa}, \citenamefont
  {Stiles},\ and\ \citenamefont {Grollier}}]{Torrejon:2017hj}%
  \BibitemOpen
  \bibfield  {author} {\bibinfo {author} {\bibfnamefont {J.}~\bibnamefont
  {Torrejon}}, \bibinfo {author} {\bibfnamefont {M.}~\bibnamefont {Riou}},
  \bibinfo {author} {\bibfnamefont {F.~A.}\ \bibnamefont {Araujo}}, \bibinfo
  {author} {\bibfnamefont {S.}~\bibnamefont {Tsunegi}}, \bibinfo {author}
  {\bibfnamefont {G.}~\bibnamefont {Khalsa}}, \bibinfo {author} {\bibfnamefont
  {D.}~\bibnamefont {Querlioz}}, \bibinfo {author} {\bibfnamefont
  {P.}~\bibnamefont {Bortolotti}}, \bibinfo {author} {\bibfnamefont
  {V.}~\bibnamefont {Cros}}, \bibinfo {author} {\bibfnamefont {K.}~\bibnamefont
  {Yakushiji}}, \bibinfo {author} {\bibfnamefont {A.}~\bibnamefont
  {Fukushima}}, \bibinfo {author} {\bibfnamefont {H.}~\bibnamefont {Kubota}},
  \bibinfo {author} {\bibfnamefont {S.}~\bibnamefont {Yuasa}}, \bibinfo
  {author} {\bibfnamefont {M.~D.}\ \bibnamefont {Stiles}}, \ and\ \bibinfo
  {author} {\bibfnamefont {J.}~\bibnamefont {Grollier}},\ }\bibfield  {title}
  {\enquote {\bibinfo {title} {{Neuromorphic computing with nanoscale
  spintronic oscillators}},}\ }\href {\doibase 10.1038/nature23011} {\bibfield
  {journal} {\bibinfo  {journal} {Nature}\ }\textbf {\bibinfo {volume} {547}},\
  \bibinfo {pages} {428--431} (\bibinfo {year} {2017})}\BibitemShut {NoStop}%
\bibitem [{\citenamefont {Romera}\ \emph {et~al.}(2018)\citenamefont {Romera},
  \citenamefont {Talatchian}, \citenamefont {Tsunegi}, \citenamefont
  {Abreu~Araujo}, \citenamefont {Cros}, \citenamefont {Bortolotti},
  \citenamefont {Trastoy}, \citenamefont {Yakushiji}, \citenamefont
  {Fukushima}, \citenamefont {Kubota}, \citenamefont {Yuasa}, \citenamefont
  {Ernoult}, \citenamefont {Vodenicarevic}, \citenamefont {Hirtzlin},
  \citenamefont {Locatelli}, \citenamefont {Querlioz},\ and\ \citenamefont
  {Grollier}}]{Romera:2018dm}%
  \BibitemOpen
  \bibfield  {author} {\bibinfo {author} {\bibfnamefont {M.}~\bibnamefont
  {Romera}}, \bibinfo {author} {\bibfnamefont {P.}~\bibnamefont {Talatchian}},
  \bibinfo {author} {\bibfnamefont {S.}~\bibnamefont {Tsunegi}}, \bibinfo
  {author} {\bibfnamefont {F.}~\bibnamefont {Abreu~Araujo}}, \bibinfo {author}
  {\bibfnamefont {V.}~\bibnamefont {Cros}}, \bibinfo {author} {\bibfnamefont
  {P.}~\bibnamefont {Bortolotti}}, \bibinfo {author} {\bibfnamefont
  {J.}~\bibnamefont {Trastoy}}, \bibinfo {author} {\bibfnamefont
  {K.}~\bibnamefont {Yakushiji}}, \bibinfo {author} {\bibfnamefont
  {A.}~\bibnamefont {Fukushima}}, \bibinfo {author} {\bibfnamefont
  {H.}~\bibnamefont {Kubota}}, \bibinfo {author} {\bibfnamefont
  {S.}~\bibnamefont {Yuasa}}, \bibinfo {author} {\bibfnamefont
  {M.}~\bibnamefont {Ernoult}}, \bibinfo {author} {\bibfnamefont
  {D.}~\bibnamefont {Vodenicarevic}}, \bibinfo {author} {\bibfnamefont
  {T.}~\bibnamefont {Hirtzlin}}, \bibinfo {author} {\bibfnamefont
  {N.}~\bibnamefont {Locatelli}}, \bibinfo {author} {\bibfnamefont
  {D.}~\bibnamefont {Querlioz}}, \ and\ \bibinfo {author} {\bibfnamefont
  {J.}~\bibnamefont {Grollier}},\ }\bibfield  {title} {\enquote {\bibinfo
  {title} {{Vowel recognition with four coupled spin-torque
  nano-oscillators}},}\ }\href {\doibase 10.1038/s41586-018-0632-y} {\bibfield
  {journal} {\bibinfo  {journal} {Nature}\ }\textbf {\bibinfo {volume} {563}},\
  \bibinfo {pages} {230--234} (\bibinfo {year} {2018})}\BibitemShut {NoStop}%
\bibitem [{\citenamefont {Riou}\ \emph {et~al.}(2019)\citenamefont {Riou},
  \citenamefont {Torrejon}, \citenamefont {Garitaine}, \citenamefont {Araujo},
  \citenamefont {Bortolotti}, \citenamefont {Cros}, \citenamefont {Tsunegi},
  \citenamefont {Yakushiji}, \citenamefont {Fukushima}, \citenamefont {Kubota},
  \citenamefont {Yuasa}, \citenamefont {Querlioz}, \citenamefont {Stiles},\
  and\ \citenamefont {Grollier}}]{Riou:2019ik}%
  \BibitemOpen
  \bibfield  {author} {\bibinfo {author} {\bibfnamefont {M.}~\bibnamefont
  {Riou}}, \bibinfo {author} {\bibfnamefont {J.}~\bibnamefont {Torrejon}},
  \bibinfo {author} {\bibfnamefont {B.}~\bibnamefont {Garitaine}}, \bibinfo
  {author} {\bibfnamefont {F.~A.}\ \bibnamefont {Araujo}}, \bibinfo {author}
  {\bibfnamefont {P.}~\bibnamefont {Bortolotti}}, \bibinfo {author}
  {\bibfnamefont {V.}~\bibnamefont {Cros}}, \bibinfo {author} {\bibfnamefont
  {S.}~\bibnamefont {Tsunegi}}, \bibinfo {author} {\bibfnamefont
  {K.}~\bibnamefont {Yakushiji}}, \bibinfo {author} {\bibfnamefont
  {A.}~\bibnamefont {Fukushima}}, \bibinfo {author} {\bibfnamefont
  {H.}~\bibnamefont {Kubota}}, \bibinfo {author} {\bibfnamefont
  {S.}~\bibnamefont {Yuasa}}, \bibinfo {author} {\bibfnamefont
  {D.}~\bibnamefont {Querlioz}}, \bibinfo {author} {\bibfnamefont {M.~D.}\
  \bibnamefont {Stiles}}, \ and\ \bibinfo {author} {\bibfnamefont
  {J.}~\bibnamefont {Grollier}},\ }\bibfield  {title} {\enquote {\bibinfo
  {title} {{Temporal Pattern Recognition with Delayed-Feedback Spin-Torque
  Nano-Oscillators}},}\ }\href {\doibase 10.1103/physrevapplied.12.024049}
  {\bibfield  {journal} {\bibinfo  {journal} {Physical Review Applied}\
  }\textbf {\bibinfo {volume} {12}},\ \bibinfo {pages} {024049} (\bibinfo
  {year} {2019})}\BibitemShut {NoStop}%
\bibitem [{\citenamefont {Tiberkevich}\ \emph {et~al.}(2014)\citenamefont
  {Tiberkevich}, \citenamefont {Khymyn}, \citenamefont {Tang},\ and\
  \citenamefont {Slavin}}]{Tiberkevich:2014eh}%
  \BibitemOpen
  \bibfield  {author} {\bibinfo {author} {\bibfnamefont {V.~S.}\ \bibnamefont
  {Tiberkevich}}, \bibinfo {author} {\bibfnamefont {R.~S.}\ \bibnamefont
  {Khymyn}}, \bibinfo {author} {\bibfnamefont {H.~X.}\ \bibnamefont {Tang}}, \
  and\ \bibinfo {author} {\bibfnamefont {A.~N.}\ \bibnamefont {Slavin}},\
  }\bibfield  {title} {\enquote {\bibinfo {title} {{Sensitivity to external
  signals and synchronization properties of a non-isochronous auto-oscillator
  with delayed feedback}},}\ }\href {\doibase 10.1038/srep03873} {\bibfield
  {journal} {\bibinfo  {journal} {Scientific Reports}\ }\textbf {\bibinfo
  {volume} {4}},\ \bibinfo {pages} {3873} (\bibinfo {year} {2014})}\BibitemShut
  {NoStop}%
\bibitem [{\citenamefont {Khalsa}, \citenamefont {Stiles},\ and\ \citenamefont
  {Grollier}(2015)}]{Khalsa:2015kn}%
  \BibitemOpen
  \bibfield  {author} {\bibinfo {author} {\bibfnamefont {G.}~\bibnamefont
  {Khalsa}}, \bibinfo {author} {\bibfnamefont {M.~D.}\ \bibnamefont {Stiles}},
  \ and\ \bibinfo {author} {\bibfnamefont {J.}~\bibnamefont {Grollier}},\
  }\bibfield  {title} {\enquote {\bibinfo {title} {{Critical current and
  linewidth reduction in spin-torque nano-oscillators by delayed
  self-injection}},}\ }\href {\doibase 10.1063/1.4922740} {\bibfield  {journal}
  {\bibinfo  {journal} {Applied Physics Letters}\ }\textbf {\bibinfo {volume}
  {106}},\ \bibinfo {pages} {242402} (\bibinfo {year} {2015})}\BibitemShut
  {NoStop}%
\bibitem [{\citenamefont {Tamaru}\ \emph {et~al.}(2016)\citenamefont {Tamaru},
  \citenamefont {Kubota}, \citenamefont {Yakushiji}, \citenamefont
  {Fukushima},\ and\ \citenamefont {Yuasa}}]{Tamaru:2016kl}%
  \BibitemOpen
  \bibfield  {author} {\bibinfo {author} {\bibfnamefont {S.}~\bibnamefont
  {Tamaru}}, \bibinfo {author} {\bibfnamefont {H.}~\bibnamefont {Kubota}},
  \bibinfo {author} {\bibfnamefont {K.}~\bibnamefont {Yakushiji}}, \bibinfo
  {author} {\bibfnamefont {A.}~\bibnamefont {Fukushima}}, \ and\ \bibinfo
  {author} {\bibfnamefont {S.}~\bibnamefont {Yuasa}},\ }\bibfield  {title}
  {\enquote {\bibinfo {title} {{Analysis of phase noise in a spin torque
  oscillator stabilized by phase locked loop}},}\ }\href {\doibase
  10.7567/apex.9.053005} {\bibfield  {journal} {\bibinfo  {journal} {Applied
  Physics Express}\ }\textbf {\bibinfo {volume} {9}},\ \bibinfo {pages}
  {053005} (\bibinfo {year} {2016})}\BibitemShut {NoStop}%
\bibitem [{\citenamefont {Tsunegi}\ \emph {et~al.}(2016)\citenamefont
  {Tsunegi}, \citenamefont {Grimaldi}, \citenamefont {Lebrun}, \citenamefont
  {Kubota}, \citenamefont {Jenkins}, \citenamefont {Yakushiji}, \citenamefont
  {Fukushima}, \citenamefont {Bortolotti}, \citenamefont {Grollier},
  \citenamefont {Yuasa},\ and\ \citenamefont {Cros}}]{Tsunegi:2016ka}%
  \BibitemOpen
  \bibfield  {author} {\bibinfo {author} {\bibfnamefont {S.}~\bibnamefont
  {Tsunegi}}, \bibinfo {author} {\bibfnamefont {E.}~\bibnamefont {Grimaldi}},
  \bibinfo {author} {\bibfnamefont {R.}~\bibnamefont {Lebrun}}, \bibinfo
  {author} {\bibfnamefont {H.}~\bibnamefont {Kubota}}, \bibinfo {author}
  {\bibfnamefont {A.~S.}\ \bibnamefont {Jenkins}}, \bibinfo {author}
  {\bibfnamefont {K.}~\bibnamefont {Yakushiji}}, \bibinfo {author}
  {\bibfnamefont {A.}~\bibnamefont {Fukushima}}, \bibinfo {author}
  {\bibfnamefont {P.}~\bibnamefont {Bortolotti}}, \bibinfo {author}
  {\bibfnamefont {J.}~\bibnamefont {Grollier}}, \bibinfo {author}
  {\bibfnamefont {S.}~\bibnamefont {Yuasa}}, \ and\ \bibinfo {author}
  {\bibfnamefont {V.}~\bibnamefont {Cros}},\ }\bibfield  {title} {\enquote
  {\bibinfo {title} {{Self-Injection Locking of a Vortex Spin Torque Oscillator
  by Delayed Feedback}},}\ }\href {\doibase 10.1038/srep26849} {\bibfield
  {journal} {\bibinfo  {journal} {Scientific Reports}\ }\textbf {\bibinfo
  {volume} {6}},\ \bibinfo {pages} {26849} (\bibinfo {year}
  {2016})}\BibitemShut {NoStop}%
\bibitem [{\citenamefont {Singh}\ \emph {et~al.}(2017)\citenamefont {Singh},
  \citenamefont {Konishi}, \citenamefont {Bhuktare}, \citenamefont {Bose},
  \citenamefont {Miwa}, \citenamefont {Fukushima}, \citenamefont {Yakushiji},
  \citenamefont {Yuasa}, \citenamefont {Kubota}, \citenamefont {Suzuki},\ and\
  \citenamefont {Tulapurkar}}]{Singh:2017gt}%
  \BibitemOpen
  \bibfield  {author} {\bibinfo {author} {\bibfnamefont {H.}~\bibnamefont
  {Singh}}, \bibinfo {author} {\bibfnamefont {K.}~\bibnamefont {Konishi}},
  \bibinfo {author} {\bibfnamefont {S.}~\bibnamefont {Bhuktare}}, \bibinfo
  {author} {\bibfnamefont {A.}~\bibnamefont {Bose}}, \bibinfo {author}
  {\bibfnamefont {S.}~\bibnamefont {Miwa}}, \bibinfo {author} {\bibfnamefont
  {A.}~\bibnamefont {Fukushima}}, \bibinfo {author} {\bibfnamefont
  {K.}~\bibnamefont {Yakushiji}}, \bibinfo {author} {\bibfnamefont
  {S.}~\bibnamefont {Yuasa}}, \bibinfo {author} {\bibfnamefont
  {H.}~\bibnamefont {Kubota}}, \bibinfo {author} {\bibfnamefont
  {Y.}~\bibnamefont {Suzuki}}, \ and\ \bibinfo {author} {\bibfnamefont {A.~A.}\
  \bibnamefont {Tulapurkar}},\ }\bibfield  {title} {\enquote {\bibinfo {title}
  {{Integer, Fractional, and Sideband Injection Locking of a Spintronic
  Feedback Nano-Oscillator to a Microwave Signal}},}\ }\href {\doibase
  10.1103/physrevapplied.8.064011} {\bibfield  {journal} {\bibinfo  {journal}
  {Physical Review Applied}\ }\textbf {\bibinfo {volume} {8}},\ \bibinfo
  {pages} {064011} (\bibinfo {year} {2017})}\BibitemShut {NoStop}%
\bibitem [{\citenamefont {Singh}\ \emph {et~al.}(2018)\citenamefont {Singh},
  \citenamefont {Bose}, \citenamefont {Bhuktare}, \citenamefont {Fukushima},
  \citenamefont {Yakushiji}, \citenamefont {Yuasa}, \citenamefont {Kubota},\
  and\ \citenamefont {Tulapurkar}}]{Singh:2018gr}%
  \BibitemOpen
  \bibfield  {author} {\bibinfo {author} {\bibfnamefont {H.}~\bibnamefont
  {Singh}}, \bibinfo {author} {\bibfnamefont {A.}~\bibnamefont {Bose}},
  \bibinfo {author} {\bibfnamefont {S.}~\bibnamefont {Bhuktare}}, \bibinfo
  {author} {\bibfnamefont {A.}~\bibnamefont {Fukushima}}, \bibinfo {author}
  {\bibfnamefont {K.}~\bibnamefont {Yakushiji}}, \bibinfo {author}
  {\bibfnamefont {S.}~\bibnamefont {Yuasa}}, \bibinfo {author} {\bibfnamefont
  {H.}~\bibnamefont {Kubota}}, \ and\ \bibinfo {author} {\bibfnamefont {A.~A.}\
  \bibnamefont {Tulapurkar}},\ }\bibfield  {title} {\enquote {\bibinfo {title}
  {{Self-Injection Locking of a Spin Torque Nano-Oscillator to Magnetic Field
  Feedback}},}\ }\href {\doibase 10.1103/physrevapplied.10.024001} {\bibfield
  {journal} {\bibinfo  {journal} {Physical Review Applied}\ }\textbf {\bibinfo
  {volume} {10}},\ \bibinfo {pages} {024001} (\bibinfo {year}
  {2018})}\BibitemShut {NoStop}%
\bibitem [{\citenamefont {Williame}\ and\ \citenamefont
  {Kim}(2020)}]{williame:2020eo}%
  \BibitemOpen
  \bibfield  {author} {\bibinfo {author} {\bibfnamefont {J.}~\bibnamefont
  {Williame}}\ and\ \bibinfo {author} {\bibfnamefont {J.-V.}\ \bibnamefont
  {Kim}},\ }\bibfield  {title} {\enquote {\bibinfo {title} {{Effects of delayed
  feedback on the power spectrum of spin-torque nano-oscillators}},}\ }\href
  {\doibase 10.1088/1361-6463/abaf26} {\bibfield  {journal} {\bibinfo
  {journal} {Journal of Physics D: Applied Physics}\ }\textbf {\bibinfo
  {volume} {53}},\ \bibinfo {pages} {495001} (\bibinfo {year}
  {2020})}\BibitemShut {NoStop}%
\bibitem [{\citenamefont {Williame}\ \emph {et~al.}(2019)\citenamefont
  {Williame}, \citenamefont {Difini~Accioly}, \citenamefont {Rontani},
  \citenamefont {Sciamanna},\ and\ \citenamefont {Kim}}]{Williame:2019hd}%
  \BibitemOpen
  \bibfield  {author} {\bibinfo {author} {\bibfnamefont {J.}~\bibnamefont
  {Williame}}, \bibinfo {author} {\bibfnamefont {A.}~\bibnamefont
  {Difini~Accioly}}, \bibinfo {author} {\bibfnamefont {D.}~\bibnamefont
  {Rontani}}, \bibinfo {author} {\bibfnamefont {M.}~\bibnamefont {Sciamanna}},
  \ and\ \bibinfo {author} {\bibfnamefont {J.-V.}\ \bibnamefont {Kim}},\
  }\bibfield  {title} {\enquote {\bibinfo {title} {{Chaotic dynamics in a
  macrospin spin-torque nano-oscillator with delayed feedback}},}\ }\href
  {\doibase 10.1063/1.5095630} {\bibfield  {journal} {\bibinfo  {journal}
  {Applied Physics Letters}\ }\textbf {\bibinfo {volume} {114}},\ \bibinfo
  {pages} {232405} (\bibinfo {year} {2019})}\BibitemShut {NoStop}%
\bibitem [{\citenamefont {Taniguchi}\ \emph {et~al.}(2019)\citenamefont
  {Taniguchi}, \citenamefont {Akashi}, \citenamefont {Notsu}, \citenamefont
  {Kimura}, \citenamefont {Tsukahara},\ and\ \citenamefont
  {Nakajima}}]{Taniguchi:2019ej}%
  \BibitemOpen
  \bibfield  {author} {\bibinfo {author} {\bibfnamefont {T.}~\bibnamefont
  {Taniguchi}}, \bibinfo {author} {\bibfnamefont {N.}~\bibnamefont {Akashi}},
  \bibinfo {author} {\bibfnamefont {H.}~\bibnamefont {Notsu}}, \bibinfo
  {author} {\bibfnamefont {M.}~\bibnamefont {Kimura}}, \bibinfo {author}
  {\bibfnamefont {H.}~\bibnamefont {Tsukahara}}, \ and\ \bibinfo {author}
  {\bibfnamefont {K.}~\bibnamefont {Nakajima}},\ }\bibfield  {title} {\enquote
  {\bibinfo {title} {{Chaos in nanomagnet via feedback current}},}\ }\href
  {\doibase 10.1103/physrevb.100.174425} {\bibfield  {journal} {\bibinfo
  {journal} {Physical Review B}\ }\textbf {\bibinfo {volume} {100}},\ \bibinfo
  {pages} {174425} (\bibinfo {year} {2019})}\BibitemShut {NoStop}%
\bibitem [{\citenamefont {Dixit}\ \emph {et~al.}(2012)\citenamefont {Dixit},
  \citenamefont {Konishi}, \citenamefont {Tomy}, \citenamefont {Suzuki},\ and\
  \citenamefont {Tulapurkar}}]{Dixit:2012ec}%
  \BibitemOpen
  \bibfield  {author} {\bibinfo {author} {\bibfnamefont {D.}~\bibnamefont
  {Dixit}}, \bibinfo {author} {\bibfnamefont {K.}~\bibnamefont {Konishi}},
  \bibinfo {author} {\bibfnamefont {C.~V.}\ \bibnamefont {Tomy}}, \bibinfo
  {author} {\bibfnamefont {Y.}~\bibnamefont {Suzuki}}, \ and\ \bibinfo {author}
  {\bibfnamefont {A.~A.}\ \bibnamefont {Tulapurkar}},\ }\bibfield  {title}
  {\enquote {\bibinfo {title} {{Spintronic oscillator based on magnetic field
  feedback}},}\ }\href {\doibase 10.1063/1.4752008} {\bibfield  {journal}
  {\bibinfo  {journal} {Applied Physics Letters}\ }\textbf {\bibinfo {volume}
  {101}},\ \bibinfo {pages} {122410} (\bibinfo {year} {2012})}\BibitemShut
  {NoStop}%
\bibitem [{\citenamefont {Kumar}\ \emph {et~al.}(2016)\citenamefont {Kumar},
  \citenamefont {Konishi}, \citenamefont {Kumar}, \citenamefont {Miwa},
  \citenamefont {Fukushima}, \citenamefont {Yakushiji}, \citenamefont {Yuasa},
  \citenamefont {Kubota}, \citenamefont {Tomy}, \citenamefont {Prabhakar},
  \citenamefont {Suzuki},\ and\ \citenamefont {Tulapurkar}}]{Kumar:2016io}%
  \BibitemOpen
  \bibfield  {author} {\bibinfo {author} {\bibfnamefont {D.}~\bibnamefont
  {Kumar}}, \bibinfo {author} {\bibfnamefont {K.}~\bibnamefont {Konishi}},
  \bibinfo {author} {\bibfnamefont {N.}~\bibnamefont {Kumar}}, \bibinfo
  {author} {\bibfnamefont {S.}~\bibnamefont {Miwa}}, \bibinfo {author}
  {\bibfnamefont {A.}~\bibnamefont {Fukushima}}, \bibinfo {author}
  {\bibfnamefont {K.}~\bibnamefont {Yakushiji}}, \bibinfo {author}
  {\bibfnamefont {S.}~\bibnamefont {Yuasa}}, \bibinfo {author} {\bibfnamefont
  {H.}~\bibnamefont {Kubota}}, \bibinfo {author} {\bibfnamefont {C.~V.}\
  \bibnamefont {Tomy}}, \bibinfo {author} {\bibfnamefont {A.}~\bibnamefont
  {Prabhakar}}, \bibinfo {author} {\bibfnamefont {Y.}~\bibnamefont {Suzuki}}, \
  and\ \bibinfo {author} {\bibfnamefont {A.}~\bibnamefont {Tulapurkar}},\
  }\bibfield  {title} {\enquote {\bibinfo {title} {{Coherent microwave
  generation by spintronic feedback oscillator}},}\ }\href {\doibase
  10.1038/srep30747} {\bibfield  {journal} {\bibinfo  {journal} {Scientific
  Reports}\ }\textbf {\bibinfo {volume} {6}},\ \bibinfo {pages} {30747}
  (\bibinfo {year} {2016})}\BibitemShut {NoStop}%
\bibitem [{\citenamefont {Bhuktare}\ \emph {et~al.}(2017)\citenamefont
  {Bhuktare}, \citenamefont {Singh}, \citenamefont {Bose},\ and\ \citenamefont
  {Tulapurkar}}]{Bhuktare:2017ja}%
  \BibitemOpen
  \bibfield  {author} {\bibinfo {author} {\bibfnamefont {S.}~\bibnamefont
  {Bhuktare}}, \bibinfo {author} {\bibfnamefont {H.}~\bibnamefont {Singh}},
  \bibinfo {author} {\bibfnamefont {A.}~\bibnamefont {Bose}}, \ and\ \bibinfo
  {author} {\bibfnamefont {A.~A.}\ \bibnamefont {Tulapurkar}},\ }\bibfield
  {title} {\enquote {\bibinfo {title} {{Spintronic Oscillator Based on
  Spin-Current Feedback Using the Spin Hall Effect}},}\ }\href {\doibase
  10.1103/physrevapplied.7.014022} {\bibfield  {journal} {\bibinfo  {journal}
  {Physical Review Applied}\ }\textbf {\bibinfo {volume} {7}},\ \bibinfo
  {pages} {014022} (\bibinfo {year} {2017})}\BibitemShut {NoStop}%
\bibitem [{\citenamefont {Appeltant}\ \emph {et~al.}(2011)\citenamefont
  {Appeltant}, \citenamefont {Soriano}, \citenamefont {Van~der Sande},
  \citenamefont {Danckaert}, \citenamefont {Massar}, \citenamefont {Dambre},
  \citenamefont {Schrauwen}, \citenamefont {Mirasso},\ and\ \citenamefont
  {Fischer}}]{Appeltant:2011jy}%
  \BibitemOpen
  \bibfield  {author} {\bibinfo {author} {\bibfnamefont {L.}~\bibnamefont
  {Appeltant}}, \bibinfo {author} {\bibfnamefont {M.~C.}\ \bibnamefont
  {Soriano}}, \bibinfo {author} {\bibfnamefont {G.}~\bibnamefont {Van~der
  Sande}}, \bibinfo {author} {\bibfnamefont {J.}~\bibnamefont {Danckaert}},
  \bibinfo {author} {\bibfnamefont {S.}~\bibnamefont {Massar}}, \bibinfo
  {author} {\bibfnamefont {J.}~\bibnamefont {Dambre}}, \bibinfo {author}
  {\bibfnamefont {B.}~\bibnamefont {Schrauwen}}, \bibinfo {author}
  {\bibfnamefont {C.~R.}\ \bibnamefont {Mirasso}}, \ and\ \bibinfo {author}
  {\bibfnamefont {I.}~\bibnamefont {Fischer}},\ }\bibfield  {title} {\enquote
  {\bibinfo {title} {{Information processing using a single dynamical node as
  complex system}},}\ }\href {\doibase 10.1038/ncomms1476} {\bibfield
  {journal} {\bibinfo  {journal} {Nature Communications}\ }\textbf {\bibinfo
  {volume} {2}},\ \bibinfo {pages} {468} (\bibinfo {year} {2011})}\BibitemShut
  {NoStop}%
\bibitem [{\citenamefont {Larger}\ \emph {et~al.}(2017)\citenamefont {Larger},
  \citenamefont {Bayl{\'o}n-Fuentes}, \citenamefont {Martinenghi},
  \citenamefont {Udaltsov}, \citenamefont {Chembo},\ and\ \citenamefont
  {Jacquot}}]{Larger:2017cx}%
  \BibitemOpen
  \bibfield  {author} {\bibinfo {author} {\bibfnamefont {L.}~\bibnamefont
  {Larger}}, \bibinfo {author} {\bibfnamefont {A.}~\bibnamefont
  {Bayl{\'o}n-Fuentes}}, \bibinfo {author} {\bibfnamefont {R.}~\bibnamefont
  {Martinenghi}}, \bibinfo {author} {\bibfnamefont {V.~S.}\ \bibnamefont
  {Udaltsov}}, \bibinfo {author} {\bibfnamefont {Y.~K.}\ \bibnamefont
  {Chembo}}, \ and\ \bibinfo {author} {\bibfnamefont {M.}~\bibnamefont
  {Jacquot}},\ }\bibfield  {title} {\enquote {\bibinfo {title} {{High-Speed
  Photonic Reservoir Computing Using a Time-Delay-Based Architecture: Million
  Words per Second Classification}},}\ }\href {\doibase
  10.1103/physrevx.7.011015} {\bibfield  {journal} {\bibinfo  {journal}
  {Physical Review X}\ }\textbf {\bibinfo {volume} {7}},\ \bibinfo {pages}
  {011015} (\bibinfo {year} {2017})}\BibitemShut {NoStop}%
\bibitem [{\citenamefont {Bourianoff}\ \emph {et~al.}(2018)\citenamefont
  {Bourianoff}, \citenamefont {Pinna}, \citenamefont {Sitte},\ and\
  \citenamefont {Everschor-Sitte}}]{Bourianoff:2018gp}%
  \BibitemOpen
  \bibfield  {author} {\bibinfo {author} {\bibfnamefont {G.}~\bibnamefont
  {Bourianoff}}, \bibinfo {author} {\bibfnamefont {D.}~\bibnamefont {Pinna}},
  \bibinfo {author} {\bibfnamefont {M.}~\bibnamefont {Sitte}}, \ and\ \bibinfo
  {author} {\bibfnamefont {K.}~\bibnamefont {Everschor-Sitte}},\ }\bibfield
  {title} {\enquote {\bibinfo {title} {{Potential implementation of reservoir
  computing models based on magnetic skyrmions}},}\ }\href {\doibase
  10.1063/1.5006918} {\bibfield  {journal} {\bibinfo  {journal} {AIP Advances}\
  }\textbf {\bibinfo {volume} {8}},\ \bibinfo {pages} {055602} (\bibinfo {year}
  {2018})}\BibitemShut {NoStop}%
\bibitem [{\citenamefont {Prychynenko}\ \emph {et~al.}(2018)\citenamefont
  {Prychynenko}, \citenamefont {Sitte}, \citenamefont {Litzius}, \citenamefont
  {Kr{\"u}ger}, \citenamefont {Bourianoff}, \citenamefont {Kl{\"a}ui},
  \citenamefont {Sinova},\ and\ \citenamefont
  {Everschor-Sitte}}]{Prychynenko:2018ii}%
  \BibitemOpen
  \bibfield  {author} {\bibinfo {author} {\bibfnamefont {D.}~\bibnamefont
  {Prychynenko}}, \bibinfo {author} {\bibfnamefont {M.}~\bibnamefont {Sitte}},
  \bibinfo {author} {\bibfnamefont {K.}~\bibnamefont {Litzius}}, \bibinfo
  {author} {\bibfnamefont {B.}~\bibnamefont {Kr{\"u}ger}}, \bibinfo {author}
  {\bibfnamefont {G.}~\bibnamefont {Bourianoff}}, \bibinfo {author}
  {\bibfnamefont {M.}~\bibnamefont {Kl{\"a}ui}}, \bibinfo {author}
  {\bibfnamefont {J.}~\bibnamefont {Sinova}}, \ and\ \bibinfo {author}
  {\bibfnamefont {K.}~\bibnamefont {Everschor-Sitte}},\ }\bibfield  {title}
  {\enquote {\bibinfo {title} {{Magnetic Skyrmion as a Nonlinear Resistive
  Element: A Potential Building Block for Reservoir Computing}},}\ }\href
  {\doibase 10.1103/physrevapplied.9.014034} {\bibfield  {journal} {\bibinfo
  {journal} {Physical Review Applied}\ }\textbf {\bibinfo {volume} {9}},\
  \bibinfo {pages} {014034} (\bibinfo {year} {2018})}\BibitemShut {NoStop}%
\bibitem [{\citenamefont {Nakane}, \citenamefont {Tanaka},\ and\ \citenamefont
  {Hirose}(2018)}]{Nakane:2018jc}%
  \BibitemOpen
  \bibfield  {author} {\bibinfo {author} {\bibfnamefont {R.}~\bibnamefont
  {Nakane}}, \bibinfo {author} {\bibfnamefont {G.}~\bibnamefont {Tanaka}}, \
  and\ \bibinfo {author} {\bibfnamefont {A.}~\bibnamefont {Hirose}},\
  }\bibfield  {title} {\enquote {\bibinfo {title} {{Reservoir Computing With
  Spin Waves Excited in a Garnet Film}},}\ }\href {\doibase
  10.1109/access.2018.2794584} {\bibfield  {journal} {\bibinfo  {journal} {IEEE
  Access}\ }\textbf {\bibinfo {volume} {6}},\ \bibinfo {pages} {4462--4469}
  (\bibinfo {year} {2018})}\BibitemShut {NoStop}%
\bibitem [{\citenamefont {Markovi{\'c}}\ \emph {et~al.}(2019)\citenamefont
  {Markovi{\'c}}, \citenamefont {Leroux}, \citenamefont {Riou}, \citenamefont
  {Araujo}, \citenamefont {Torrejon}, \citenamefont {Querlioz}, \citenamefont
  {Fukushima}, \citenamefont {Yuasa}, \citenamefont {Trastoy}, \citenamefont
  {Bortolotti},\ and\ \citenamefont {Grollier}}]{Markovic:2019dz}%
  \BibitemOpen
  \bibfield  {author} {\bibinfo {author} {\bibfnamefont {D.}~\bibnamefont
  {Markovi{\'c}}}, \bibinfo {author} {\bibfnamefont {N.}~\bibnamefont
  {Leroux}}, \bibinfo {author} {\bibfnamefont {M.}~\bibnamefont {Riou}},
  \bibinfo {author} {\bibfnamefont {F.~A.}\ \bibnamefont {Araujo}}, \bibinfo
  {author} {\bibfnamefont {J.}~\bibnamefont {Torrejon}}, \bibinfo {author}
  {\bibfnamefont {D.}~\bibnamefont {Querlioz}}, \bibinfo {author}
  {\bibfnamefont {A.}~\bibnamefont {Fukushima}}, \bibinfo {author}
  {\bibfnamefont {S.}~\bibnamefont {Yuasa}}, \bibinfo {author} {\bibfnamefont
  {J.}~\bibnamefont {Trastoy}}, \bibinfo {author} {\bibfnamefont
  {P.}~\bibnamefont {Bortolotti}}, \ and\ \bibinfo {author} {\bibfnamefont
  {J.}~\bibnamefont {Grollier}},\ }\bibfield  {title} {\enquote {\bibinfo
  {title} {{Reservoir computing with the frequency, phase, and amplitude of
  spin-torque nano-oscillators}},}\ }\href {\doibase 10.1063/1.5079305}
  {\bibfield  {journal} {\bibinfo  {journal} {Applied Physics Letters}\
  }\textbf {\bibinfo {volume} {114}},\ \bibinfo {pages} {012409} (\bibinfo
  {year} {2019})}\BibitemShut {NoStop}%
\bibitem [{\citenamefont {Araujo}\ \emph {et~al.}(2020)\citenamefont {Araujo},
  \citenamefont {Riou}, \citenamefont {Torrejon}, \citenamefont {Tsunegi},
  \citenamefont {Querlioz}, \citenamefont {Yakushiji}, \citenamefont
  {Fukushima}, \citenamefont {Kubota}, \citenamefont {Yuasa}, \citenamefont
  {Stiles},\ and\ \citenamefont {Grollier}}]{Araujo:2020ro}%
  \BibitemOpen
  \bibfield  {author} {\bibinfo {author} {\bibfnamefont {F.~A.}\ \bibnamefont
  {Araujo}}, \bibinfo {author} {\bibfnamefont {M.}~\bibnamefont {Riou}},
  \bibinfo {author} {\bibfnamefont {J.}~\bibnamefont {Torrejon}}, \bibinfo
  {author} {\bibfnamefont {S.}~\bibnamefont {Tsunegi}}, \bibinfo {author}
  {\bibfnamefont {D.}~\bibnamefont {Querlioz}}, \bibinfo {author}
  {\bibfnamefont {K.}~\bibnamefont {Yakushiji}}, \bibinfo {author}
  {\bibfnamefont {A.}~\bibnamefont {Fukushima}}, \bibinfo {author}
  {\bibfnamefont {H.}~\bibnamefont {Kubota}}, \bibinfo {author} {\bibfnamefont
  {S.}~\bibnamefont {Yuasa}}, \bibinfo {author} {\bibfnamefont {M.~D.}\
  \bibnamefont {Stiles}}, \ and\ \bibinfo {author} {\bibfnamefont
  {J.}~\bibnamefont {Grollier}},\ }\bibfield  {title} {\enquote {\bibinfo
  {title} {{Role of non-linear data processing on speech recognition task in
  the framework of reservoir computing}},}\ }\href {\doibase
  10.1038/s41598-019-56991-x} {\bibfield  {journal} {\bibinfo  {journal}
  {Scientific Reports}\ }\textbf {\bibinfo {volume} {10}},\ \bibinfo {pages}
  {328} (\bibinfo {year} {2020})}\BibitemShut {NoStop}%
\bibitem [{\citenamefont {Yamaguchi}\ \emph {et~al.}(2020)\citenamefont
  {Yamaguchi}, \citenamefont {Akashi}, \citenamefont {Tsunegi}, \citenamefont
  {Kubota}, \citenamefont {Nakajima},\ and\ \citenamefont
  {Taniguchi}}]{Yamaguchi:2020hg}%
  \BibitemOpen
  \bibfield  {author} {\bibinfo {author} {\bibfnamefont {T.}~\bibnamefont
  {Yamaguchi}}, \bibinfo {author} {\bibfnamefont {N.}~\bibnamefont {Akashi}},
  \bibinfo {author} {\bibfnamefont {S.}~\bibnamefont {Tsunegi}}, \bibinfo
  {author} {\bibfnamefont {H.}~\bibnamefont {Kubota}}, \bibinfo {author}
  {\bibfnamefont {K.}~\bibnamefont {Nakajima}}, \ and\ \bibinfo {author}
  {\bibfnamefont {T.}~\bibnamefont {Taniguchi}},\ }\bibfield  {title} {\enquote
  {\bibinfo {title} {{Periodic structure of memory function in spintronics
  reservoir with feedback current}},}\ }\href {\doibase
  10.1103/physrevresearch.2.023389} {\bibfield  {journal} {\bibinfo  {journal}
  {Physical Review Research}\ }\textbf {\bibinfo {volume} {2}},\ \bibinfo
  {pages} {023389} (\bibinfo {year} {2020})}\BibitemShut {NoStop}%
\bibitem [{\citenamefont {Pinna}, \citenamefont {Bourianoff},\ and\
  \citenamefont {Everschor-Sitte}(2020)}]{Pinna:2020rc}%
  \BibitemOpen
  \bibfield  {author} {\bibinfo {author} {\bibfnamefont {D.}~\bibnamefont
  {Pinna}}, \bibinfo {author} {\bibfnamefont {G.}~\bibnamefont {Bourianoff}}, \
  and\ \bibinfo {author} {\bibfnamefont {K.}~\bibnamefont {Everschor-Sitte}},\
  }\bibfield  {title} {\enquote {\bibinfo {title} {{Reservoir Computing with
  Random Skyrmion Textures}},}\ }\href {\doibase
  10.1103/physrevapplied.14.054020} {\bibfield  {journal} {\bibinfo  {journal}
  {Physical Review Applied}\ }\textbf {\bibinfo {volume} {14}},\ \bibinfo
  {pages} {054020} (\bibinfo {year} {2020})}\BibitemShut {NoStop}%
\bibitem [{\citenamefont {Mackey}\ and\ \citenamefont
  {Glass}(1977)}]{Mackey:1977dv}%
  \BibitemOpen
  \bibfield  {author} {\bibinfo {author} {\bibfnamefont {M.~C.}\ \bibnamefont
  {Mackey}}\ and\ \bibinfo {author} {\bibfnamefont {L.}~\bibnamefont {Glass}},\
  }\bibfield  {title} {\enquote {\bibinfo {title} {{Oscillation and chaos in
  physiological control systems}},}\ }\href {\doibase 10.1126/science.267326}
  {\bibfield  {journal} {\bibinfo  {journal} {Science}\ }\textbf {\bibinfo
  {volume} {197}},\ \bibinfo {pages} {287--289} (\bibinfo {year}
  {1977})}\BibitemShut {NoStop}%
\bibitem [{\citenamefont {Vansteenkiste}\ \emph {et~al.}(2014)\citenamefont
  {Vansteenkiste}, \citenamefont {Leliaert}, \citenamefont {Dvornik},
  \citenamefont {Helsen}, \citenamefont {Garc{\'\i}a-S{\'a}nchez},\ and\
  \citenamefont {Van~Waeyenberge}}]{Vansteenkiste:2014et}%
  \BibitemOpen
  \bibfield  {author} {\bibinfo {author} {\bibfnamefont {A.}~\bibnamefont
  {Vansteenkiste}}, \bibinfo {author} {\bibfnamefont {J.}~\bibnamefont
  {Leliaert}}, \bibinfo {author} {\bibfnamefont {M.}~\bibnamefont {Dvornik}},
  \bibinfo {author} {\bibfnamefont {M.}~\bibnamefont {Helsen}}, \bibinfo
  {author} {\bibfnamefont {F.}~\bibnamefont {Garc{\'\i}a-S{\'a}nchez}}, \ and\
  \bibinfo {author} {\bibfnamefont {B.}~\bibnamefont {Van~Waeyenberge}},\
  }\bibfield  {title} {\enquote {\bibinfo {title} {{The design and verification
  of MuMax3}},}\ }\href {\doibase 10.1063/1.4899186} {\bibfield  {journal}
  {\bibinfo  {journal} {AIP Advances}\ }\textbf {\bibinfo {volume} {4}},\
  \bibinfo {pages} {107133} (\bibinfo {year} {2014})}\BibitemShut {NoStop}%
\end{thebibliography}%

\end{document}